\documentclass[12pt]{article}

\usepackage[centertags]{amsmath}
\usepackage{amsfonts}
\usepackage{amssymb}
\usepackage{amsthm}
\usepackage{newlfont}
\usepackage{epsfig}
\usepackage{amscd}

\newcommand{\RR}{{\mathbb R}}

\newcommand{\beq}{\begin{equation}}
\newcommand{\eeq}{\end{equation}}
\newcommand{\ba}{\begin{array}}
\newcommand{\ea}{\end{array}}
\newcommand{\bea}{\begin{eqnarray}}
\newcommand{\eea}{\end{eqnarray}}

\begin{document}

\begin{center}
{\large \sc \bf Inverse Scattering Problem for Vector Fields} 

\vskip 5pt
{\large \sc \bf and the Cauchy Problem for the Heavenly Equation} 

\vskip 20pt

{\large  S. V. Manakov$^{1,\S}$ and P. M. Santini$^{2,\S}$}

\vskip 20pt

{\it 
$^1$ Landau Institute for Theoretical Physics, Moscow, Russia

\smallskip

$^2$ Dipartimento di Fisica, Universit\`a di Roma "La Sapienza", and \\
Istituto Nazionale di Fisica Nucleare, Sezione di Roma 1 \\
Piazz.le Aldo Moro 2, I-00185 Roma, Italy}

\bigskip

$^{\S}$e-mail:  {\tt manakov@itp.ac.ru, paolo.santini@roma1.infn.it}

\bigskip

{\today}

\end{center}

\begin{abstract}
We solve the inverse scattering problem for multidimensional vector fields 
and we use this result to construct the formal solution of the Cauchy problem for  
the second heavenly equation of Plebanski, a scalar nonlinear partial differential equation in four  
dimensions relevant in General Relativity, which arises from the commutation of multidimensional 
Hamiltonian vector fields.  
\end{abstract}
\section{Introduction}
In this paper we solve the inverse scattering problem for 
multidimensional vector fields and we use this result to construct  
the formal solution of the Cauchy problem for the real second heavenly equation
\beq
\label{heavenly1}
\theta_{tx}-\theta_{zy}+\theta_{xx}\theta_{yy}-\theta^2_{xy}=0,~~~
\theta=\theta(x,y,z,t)\in\RR,~~~~~x,y,z,t\in\RR,
\eeq 
where subscripts denote partial derivatives.  

This scalar nonlinear partial differential equation (PDE) in $4$ independent variables 
$x,y,z,t$, introduced in \cite{Pleb} by Plebanski, describes the Einstein field equations that 
govern self-dual gravitational fields. Its 
$2$-dimensional reduction $\theta_{z}=\theta_{t}=0$ is the Monge-Amp\`ere 
equation, relevant in Differential Geometry. As we shall see in the following, 
the heavenly equation plays also a distinguished role in the theory of commuting, 2-dimensional, Hamiltonian 
dynamical systems. 
 
The heavenly equation, together with the 
equations for the self-dual Yang-Mills (SDYM) fields \cite{YM}, are perhaps the most distinguished 
examples of nonlinear PDEs in more than three independent variables arising as commutativity 
conditions of linear operators \cite{ZS}, and therefore amenable, in principle, 
to exact treatments based on the spectral theory of those operators \cite{ZMNP},\cite{AC}. If the SDYM equations 
are considered on an abstract Lie algebra, then the heavenly equation can actually be interpreted as a distinguished 
realization of the SDYM equations, corresponding to the Lie algebra of divergence free vector fields independent of 
the SDYM coordinates \cite{CMN}.  Equation (\ref{heavenly1}) has been investigated within the twistor approach  
in \cite{DuMa},\cite{DMT}. A bi-Hamiltonian formulation and a hodograph transformation for (\ref{heavenly1}) have been 
recently constructed in \cite{NNS} and \cite{MM}; a nonlinear $\bar\partial$-dressing and a generating equation for its 
hierarchy can be found in \cite{BK}. 

The present paper is a simplified version of the manuscript \cite{MS}. 

\section{Integrable PDEs in arbitrary dimensions and the heavenly equation}

It is known that the commutation of multidimensional vector fields  
leads to nonlinear first order multidimensional PDEs (see, f.i., \cite{ZS}). In this spirit, 
we derive now a basic class of integrable nonlinear PDEs in arbitrary 
dimensions possessing, as distinguished reduction, the heavenly equation (\ref{heavenly1}). 

Consider the following pair of vector fields
\beq
\label{L1L2}
\hat L_i=\partial_{t_i}+\lambda\partial_{z_i}+
\sum\limits_{k=1}^Nu^k_i\partial_{x_k}=\partial_{t_i}+\lambda\partial_{z_i}+\vec u_i\cdot\nabla_{\vec x},~~~~i=1,2 
\eeq  
where $\partial_x$ denotes partial differentiation with respect to the generic variable $x$, $\vec x=(x_1,..,x_N)$,  
$\nabla_{\vec x}=(\partial_{x_1},..,\partial_{x_N})$, $\vec u_i=(u^1_i,..,u^N_i),~i=1,2$, 
$\lambda$ is a complex parameter and the vector coefficients $\vec u_i$ depend on the 
independent variables $t_i,z_i,x_k$, $i=1,2,~k=1,..,N$, but not  
on $\lambda$. The existence of a common eigenfunction $f$ for the operators $\hat L_1$ and $\hat L_2$:
\beq
\label{Lax1}
\hat L_1f=\hat L_2f=0,
\eeq
implies their commutation, $\forall\lambda$:
\beq
[\hat L_1,\hat L_2]=0,
\eeq  
which is equivalent to the following determined system of $2N$ first order quasi-linear PDEs in $(4+N)$ dimensions 
for the $2N$ fields $\vec u_1,\vec u_2$:
\beq
\label{quasilin-u}
\ba{l}
\vec {u_1}_{z_2}=\vec {u_2}_{z_1}, \\
\vec {u_1}_{t_2}-\vec {u_2}_{t_1}+
\left(\vec u_2\cdot\nabla_{\vec x}\right)\vec u_1-\left(\vec u_1\cdot\nabla_{\vec x}\right)\vec u_2=\vec 0.
\ea
\eeq
Parametrizing the first set of equations in terms of the vector potential $\vec U$ 
\beq
\vec u_i=\vec U_{z_i},~~~~~~i=1,2,
\eeq
one obtains the following determined system of $N$ nonlinear PDEs for the $N$ dependent variables $\vec U$ in $(4+N)$ dimensions:
\beq
\label{quasilin-U}
\vec U_{t_1z_2}-\vec U_{t_2z_1}+
\left(\vec U_{z_1}\cdot\nabla_{\vec x}\right)\vec U_{z_2}-\left(\vec U_{z_2}\cdot\nabla_{\vec x}\right)\vec U_{z_1}=\vec 0.
\eeq

This system admits a natural reduction; indeed, applying the operator $\nabla_{\vec x}\cdot$ to equations 
(\ref{quasilin-U}), one obtains
\beq
\left[{\partial}_{t_1}{\partial}_{z_2}-{\partial}_{t_2}{\partial}_{z_1}+
\left(\vec U_{z_1}\cdot\nabla_{\vec x}\right){\partial}_{z_2}-
\left(\vec U_{z_2}\cdot\nabla_{\vec x}\right){\partial}_{z_1}\right]\left(\nabla_{\vec x}\cdot\vec U\right)=0,
\eeq
from which one infers that the condition
\beq
\label{constraint-U}
\nabla_{\vec x}\cdot\vec U=0
\eeq
is an admissible reduction for equation (\ref{quasilin-U}), implying that the condition of zero-divergence: 
\beq
\label{red-u}
\nabla_{\vec x}\cdot\vec u_i=0,~~~i=1,2
\eeq
is an admissible constraint for the vectors $\vec u_i,~i=1,2$. 

\vskip 5pt
From now on, we concentrate our attention on the following important example:
\beq
\label{N=2}
N=2,~~~z_i=x_i,~~~i=1,2
\eeq
and we make the following change of notation for the remaining 4 variables:
\beq
\label{xyzt}
t_1=z,~t_2=t,~~x_1=x,~~x_2=y.
\eeq
Then the system (\ref{quasilin-U}) reduces to the following determined system of two PDEs 
in 4 dimensions:
\beq
\label{quasilin-U-bis}
\ba{l}
\vec U_{tx}-\vec U_{zy}+\left(\vec U_{y}\cdot\nabla_{\vec x}\right)\vec U_{x}-
\left(\vec U_{x}\cdot\nabla_{\vec x}\right)\vec U_{y}=\vec 0, \\
\vec U\in\RR^2,~~\vec x=(x,y),~~\nabla_{\vec x}=(\partial_x,\partial_y),
\ea
\eeq
corresponding to the Lax pair:
\beq
\label{L1L2bis}
\ba{l}
\hat L_1=\partial_{z}+\lambda\partial_{x}+\vec u_1\cdot\nabla_{\vec x}, ~~~~~\vec u_1=\vec U_x, \\
\hat L_2=\partial_{t}+\lambda\partial_{y}+\vec u_2\cdot\nabla_{\vec x},~~~~~\vec u_2=\vec U_y.
\ea
\eeq
In this case, the zero-divergence reduction (\ref{red-u}) makes the two vector fields \break 
$\vec u_i\cdot\nabla_{\vec x}$ 
Hamiltonian, allowing for the introduction of two Hamiltonians $H_i,~i=1,2$ such that:
\beq
\vec u_i=({H_i}_y,-{H_i}_x),\;\;\;i=1,2
\eeq
which, due to (\ref{quasilin-u}a), are parametrized by a single potential $\theta$:
\beq
\label{def-theta}
\ba{l}
H_i=\theta_{x_i},~~~~~\vec U=(\theta_y,-\theta_x),\\
\vec u_1=(\theta_{xy},-\theta_{xx}),~~~~\vec u_2=(\theta_{yy},-\theta_{xy}).
\ea
\eeq
Then the compatible linear problems (\ref{Lax1}),(\ref{L1L2bis}) can be written down as Hamilton equations 
with respect to the times $z,t$:
\beq
\label{HamEqu}
\ba{l}
f_{z}=\{H_1+\lambda y,f\}_{\vec x}, \\
f_{t}=\{H_2-\lambda x,f\}_{\vec x}, 
\ea
\eeq
where $\{\cdot,\cdot\}_{\vec x}$ is the Poisson bracket with respect to the variables $x,y$:
\beq
\label{PB}
\{f,g\}_{\vec x}=f_{x}g_{y}-f_{y}g_{x},
\eeq
and the nonlinear system (\ref{quasilin-U-bis}) reduces to the heavenly equation 
in Hamiltonian form  
\beq
\label{heavenly2}
\theta_{tx}-\theta_{zy}+\{\theta_{x},\theta_{y}\}_{\vec x}=\mbox{constant},~~~
\eeq 
equivalent to (\ref{heavenly1}) after choosing the constant to be zero.
\section{IST for the nonlinear PDEs (\ref{quasilin-U-bis}) and (\ref{heavenly1})}
Since the Lax pair (\ref{L1L2bis}) is made of vector fields, Hamiltonian in the heavenly reduction (\ref{red-u}),(\ref{N=2}), 
the eigenfunctions satisfy the following basic properties, which will introduce important novelties in the 
Inverse Scattering Transform (IST).

\vskip 2pt
\noindent
1) {\it The space of eigenfunctions is a ring}:  if $f_1,~f_2$ are two solutions of the Lax pair 
(\ref{L1L2bis}), then an arbitrary differentiable function $F(f_1,f_2)$ of them is a solution of (\ref{L1L2bis}).   

\vskip 2pt
\noindent
2) {\it In the heavenly (Hamiltonian) reduction (\ref{red-u}),(\ref{N=2}), the space of eigenfunctions is also a Lie algebra, 
whose Lie bracket is 
the natural Poisson bracket (\ref{PB})}: if $f_1,~f_2$ are two solutions of the Lax pair (\ref{L1L2bis}), 
then their Poisson bracket $\{f_1,f_2\}_{(x,y)}$ is also a solution of (\ref{L1L2bis}).

Now we consider the Cauchy problem for the system (\ref{quasilin-U-bis}) and for its heavenly reduction (\ref{heavenly1}),    
within the class of rapidly decreasing real potentials $u^j_i$:
\beq
\label{localization}
\ba{l}
u^j_i\to~0,~~(x^2+y^2+z^2)\to\infty, \\
u^j_i\in\RR,~~~(x,y,z)\in\RR^3,~~t>0,
\ea
\eeq
interpreting $t$ as time and the other three variables $x,y,z$ as space variables.  

To solve such a Cauchy problem by the IST method (see, f.i., \cite{ZMNP},\cite{AC}), we construct the 
IST for the operator $\hat L_1$ in (\ref{L1L2bis}a), within the class of rapidly decreasing real potentials, 
interpreting the operator $\hat L_2$ in (\ref{L1L2bis}b) as the time operator. The formalism presented here can 
be generalized to the vector fields (\ref{L1L2}) in a straightforward way. 
\subsection{Basic eigenfunctions}
The localization (\ref{localization}) of the 
vector potential $\vec u_1$ implies that, if $f$ is a solution of $\hat L_1 f=0$, then 
\beq
\label{asymptf}
\ba{l}
f(\vec x,z,\lambda)\to f_{\pm}(\vec \xi,\lambda),\;\;z\to\pm\infty, \\
\vec \xi:=\vec x-(\lambda,0)z;
\ea
\eeq
i.e., asymptotically, $f$ is an arbitrary function of $(x-\lambda z)$, $y$ and $\lambda$.

A central role in the theory is played by the real Jost eigenfunctions $\varphi_{1,2}(\vec x,z,\lambda)$, 
the solutions of $\hat L_1\varphi_{1,2}=0$ uniquely defined by the asymptotics 
\beq
\label{def-varphi}
\varphi_1(\vec x,z,\lambda)\to x-\lambda z\equiv \xi,~~~~\varphi_2(\vec x,z,\lambda)\to y,~~~z\to -\infty.
\eeq
In this paper we often use the compact vector notation: $\vec f=(f_1,f_2)^T$. Then:
\beq
\label{def-vec-varphi}
\vec\varphi(\vec x,z,\lambda)\equiv \left(
\ba{l}
\varphi_1(\vec x,z,\lambda) \\
\varphi_2(\vec x,z,\lambda)
\ea
\right) \to \left(
\ba{l}
\xi \\
y
\ea
\right)\equiv \vec \xi,~~z\to -\infty.
\eeq 
The Jost eigenvector $\vec\varphi$ is equivalently characterized by the integral equation
\beq
\label{phi}
\ba{l}
\vec\varphi(\vec x,z,\lambda)+ \\
\int_{\RR^3}d{\vec x}'dz'G(\vec x-\vec x',z-z';\lambda)\left(\vec u_1(\vec x',z')\cdot\nabla_{\vec x'}\right)
\vec\varphi(\vec x',z',\lambda)=\vec\xi,
\ea
\eeq
in terms of the Jost Green's function
\beq
G(\vec x,z;\lambda)=\theta(z)\delta(x-\lambda z)\delta(y).
\eeq

A crucial role in the IST for the vector field $\hat L_1$ is also played by the   
analytic eigenfunctions $\vec\psi_{\pm}(\vec x,z,\lambda)$, the solutions of $\hat L_1\vec\psi_{\pm}=\vec 0$ 
satisfying the integral equations 
\beq
\label{psi}
\ba{l}
\vec\psi_{\pm}(\vec x,z,\lambda)+
\int_{\RR^3}d{\vec x}'dz'G_{\pm}(\vec x-\vec x',z-z';\lambda)\left(\vec u_1(\vec x',z')\cdot\nabla_{\vec x'}\right)
\vec\psi_{\pm}(\vec x',z',\lambda)=\vec \xi,
\ea
\eeq
where $G_{\pm}$ are the analytic Green's functions
\beq
\label{Green_analytic}
G_{\pm}(\vec x,z;\lambda)=\pm\frac{\delta(y)}{2\pi i[x-(\lambda\pm i\epsilon) z]}.
\eeq
The analyticity properties of $G_{\pm}(\vec x,z,\lambda)$ in the complex $\lambda$ - plane 
imply that $\vec\psi_{+}(\vec x,z,\lambda)$ and $\vec\psi_{-}(\vec x,z,\lambda)$ are 
analytic, respectively, in the upper and lower halves of the complex $\lambda$ - plane, with 
the following asymptotics, for large $\lambda$:
\beq
\label{asympt_psi}
\vec\psi_{\pm}(\vec x,z,\lambda)=\vec \xi+\frac{\vec Q_{\pm}(\vec x,z)}{\lambda}+O(\lambda^{-2}).
~~~|\lambda|>>1,
\eeq  
where:
\beq
\label{def-Q+-}
{\vec Q}_{\pm}(\vec x,z)=\pm P\int_{\RR^2}\frac{dx'dz'}{2\pi i(z-z')}\vec u_1(x',y,z')-
\frac{1}{2}\left(\int\limits_{-\infty}^x-\int\limits_{x}^{\infty}\right)dx'\vec u_1(x',y,z),
\eeq
entailing that
\beq
\label{u-Q}
\vec u_1(\vec x,z)=-{\vec Q}_{\pm x}(\vec x,z).
\eeq

It is important to remark that the analytic Green's functions (\ref{Green_analytic}) exhibit the following 
asymptotics for $z\to\pm\infty$:
\beq
\ba{l}
G_{\pm}(\vec x,z;\lambda)\to\pm\frac{\delta(y)}{2\pi i[\xi \mp i\epsilon]},\;\;z\to +\infty, \\
G_{\pm}(\vec x,z;\lambda)\to\pm\frac{\delta(y)}{2\pi i[\xi \pm i\epsilon]},\;\;z\to -\infty,
\ea
\eeq
It follows that the $z=+\infty$ asymptotics of $\vec\psi_{+}$ and $\vec\psi_{-}$ are analytic 
respectively in the lower and upper halves of the complex plane $\xi$, while the $z=-\infty$ asymptotics 
of $\vec\psi_{+}$ and $\vec\psi_{-}$ are analytic respectively in the upper and lower halves of the complex plane $\xi$. 
This mechanism was first observed in \cite{MZ}.

\subsection{Spectral data}
The $z=+\infty$ limit of $\vec\varphi$ defines the scattering vector $\vec\sigma$ of $\hat L_1$:
\beq
\label{def-S}
\displaystyle\lim_{z\to +\infty}\vec\varphi(\vec x,z;\lambda) \equiv 
\vec{\cal S}(\vec \xi,\lambda)=\vec \xi+\vec\sigma(\vec \xi,\lambda).
\eeq

The direct problem is the mapping from the real vector potential $\vec u_1$, 
function of the three real variables $(\vec x,z)$, to the real scattering vector $\vec\sigma$, function of 
the three real variables $(\vec\xi,\lambda)$.  
Then the counting is consistent. The impact of 
the Heavenly constraint (\ref{red-u}),(\ref{N=2}) on the spectral data will be discussed in Section 3.5. 

The Jost solutions $\varphi_{1,2}$ form, together with the constant eigenfunction $\lambda$, 
 a basis in the space of the eigenfunctions of $\hat L_1$ (which is a ring).  
The representation of the analytic eigenfunctions $\vec\psi_{\pm}$ in terms of $\vec\varphi$ yields:
\beq
\label{varphi-psi}
\vec\psi_{\pm}(\vec x,z,\lambda)=\vec{\cal K}_{\pm}\left(\vec\varphi(\vec x,z,\lambda),\lambda\right)=
\vec\varphi(\vec x,z,\lambda)+\vec\chi_{\pm}\left(\vec\varphi(\vec x,z,\lambda),\lambda\right),
\eeq
and this formula defines the spectral data $\vec\chi_{\pm}$. Since the $z\to -\infty$ limit of (\ref{varphi-psi}) reads:
\beq
\label{lim-varphi-psi}
\displaystyle\lim_{z\to -\infty}\vec\psi_{\pm}-\vec \xi=\vec\chi_{\pm}(\vec \xi,\lambda),
\eeq
the above analyticity properties of the LHS of (\ref{lim-varphi-psi}) in the complex 
$\xi$ - plane imply that $\vec\chi_{+}(\vec \xi,\lambda)$ and $\vec\chi_{-}(\vec \xi,\lambda)$ are analytic 
respectively in the upper and lower 
halves  of the complex plane $\xi$, decaying at $\xi\sim\infty$ like $O(\xi^{-1})$. Therefore their Fourier transforms 
$\tilde{\vec\chi}_{+}(\vec \omega,\lambda)$ and $\tilde{\vec\chi}_{-}(\vec \omega,\lambda)$ have support respectively 
on the positive and negative real $\omega_1$ semi-axes. 

The spectral data $\vec\chi_{\pm}$ can be constructed from the scattering vector $\vec\sigma$ 
through the solution of the following linear integral equations 
\beq
\label{Fourier-varphi-psi}
\ba{l}
\tilde{\vec\chi}_+(\vec\omega,\lambda)+\theta(\omega_1)\left(\tilde{\vec\sigma}(\vec\omega,\lambda)+
\int_{\RR^2}d\vec\eta ~\tilde{\vec\chi}_+(\vec\eta,\lambda)Q(\vec\eta,\vec\omega,\lambda)\right)=\vec 0,  \\
\tilde{\vec\chi}_-(\vec\omega,\lambda)+\theta(-\omega_1)\left(\tilde{\vec\sigma}(\vec\omega,\lambda)+
\int_{\RR^2}d\vec\eta ~\tilde{\vec\chi}_-(\vec\eta,\lambda)Q(\vec\eta,\vec\omega,\lambda)\right)=\vec 0,
\ea
\eeq 
involving the Fourier transforms $\tilde{\vec\sigma}$ and $\tilde{\vec\chi}_{\pm}$ of $\vec\sigma$ and ${\vec\chi}_{\pm}$:
\beq
\label{Fourier-sigma}
\tilde{\vec\sigma}(\vec\omega,\lambda)=\int_{\RR^2}d\vec \xi\vec\sigma(\vec \xi,\lambda)e^{-i\vec\omega\cdot\vec \xi},~~~
\tilde{\vec\chi}_{\pm}(\vec\omega,\lambda)=\int_{\RR^2}d\vec \xi{\vec\chi}_{\pm}(\vec \xi,\lambda)e^{-i\vec\omega\cdot\vec \xi}
\eeq
and the kernel:
\beq
\label{def-Q}
Q(\vec\eta,\vec\omega,\lambda)=\int_{\RR^2}\frac{d\vec \xi}{(2\pi)^2}e^{i(\vec\eta-\vec\omega)\cdot\vec \xi}
[e^{i\vec\eta\cdot\vec\sigma(\vec \xi,\lambda)}-1].
\eeq 
To prove this result, one first evaluates (\ref{varphi-psi}) at $z=+\infty$, obtaining 
\beq
\label{+lim-varphi-psi}
\displaystyle\lim_{z\to \infty}\vec\psi_{\pm}-\vec \xi=\vec\sigma(\vec \xi,\lambda)+
\vec\chi_{\pm}(\vec \xi+\vec\sigma(\vec \xi,\lambda),\lambda).
\eeq 
Applying the integral operator $\int_{\RR^2}d\vec \xi e^{-i\vec\omega\cdot\vec \xi}\cdot$ 
for $\omega_1>0$ and $\omega_1<0$ respectively to equations (\ref{+lim-varphi-psi})$_{+}$ and 
(\ref{+lim-varphi-psi})$_{-}$, using the above analyticity properties  
and the Fourier representations of ${\vec\chi}_{\pm}$ and $\vec\sigma$, one obtains equations 
(\ref{Fourier-varphi-psi}).

We end this section remarking that the reality of the potentials: $\vec u_1\in\RR^2$ implies that, for $\lambda\in\RR$,  
$\overline{\vec\varphi}=\vec\varphi$, $\overline{\vec\psi}_+=\vec\psi_-$; consequently: 
$\overline{\vec\sigma}=\vec\sigma$, $\overline{\vec\chi}_+=\vec\chi_-$. 
\subsection{Inverse Problem}

An inverse problem can be constructed from equations (\ref{varphi-psi}). 
Subtracting $\vec \xi$ from equations (\ref{varphi-psi})$_{-}$ and (\ref{varphi-psi})$_{+}$,  
applying respectively the analyticity projectors $\hat P_{+}$ and $\hat P_{-}$: 
\beq
\hat P_{\pm}\equiv \pm\frac{1}{2\pi i}\int_{\RR}\frac{d\lambda'}{\lambda'-(\lambda\pm i\epsilon)}. 
\eeq
and adding up the resulting equations, one obtains the following 
nonlinear integral equation for the Jost eigenfunction $\vec\varphi$:
\beq
\label{varphi-int-equ2}
\ba{l}
\vec\varphi(\vec x,z,\lambda)+\frac{1}{2\pi i}\int_{\RR}\frac{d\lambda'}{\lambda'-(\lambda+i\epsilon)}
\vec\chi_-(\vec\varphi(\vec x,z,\lambda'),\lambda') - \\
\frac{1}{2\pi i}\int_{\RR}\frac{d\lambda'}{\lambda'-(\lambda-i\epsilon)}\vec\chi_+(\vec\varphi(\vec x,z,\lambda'),\lambda')=
\vec\xi. 
\ea
\eeq
Given the spectral data ${\vec\chi}_{\pm}$, one reconstructs the eigenfunction $\vec\varphi$ from 
(\ref{varphi-int-equ2}), the analytic eigenfunctions from (\ref{varphi-psi}), 
and $\vec u_1$ from equation (\ref{u-Q}). This inversion procedure was first introduced in \cite{Manakov1}.

\subsection{$t$-evolution of the spectral data}
As the potentials ${\vec u}_{1,2}$ evolve in time according to equation (\ref{quasilin-U-bis}), the $t$-dependence of 
the spectral data $\vec\sigma$ and $\vec\chi_{\pm}$,  
defined in (\ref{def-S}) and (\ref{varphi-psi}), is described by the equation:
\beq
\label{t-dep-sigma-chi}
\ba{l}
\vec\sigma(\vec\xi,\lambda,t)=\vec\sigma(\vec\xi-(0,\lambda)t,\lambda,0), \\
\vec\chi_{\pm}(\vec\xi,\lambda,t)=\vec\chi_{\pm}(\vec\xi-(0,\lambda)t,\lambda,0).
\ea
\eeq
To prove it, we first observe that 
\beq
\label{def-phi}
\phi_1(\vec x,z,\lambda,t)\equiv \varphi_1(\vec x,z,\lambda,t), ~~
\phi_2(\vec x,z,\lambda,t)\equiv \varphi_2(\vec x,z,\lambda,t)-\lambda t
\eeq
are, together with $\lambda$, a basis of   
common Jost eigenfunctions of $\hat L_1$ and $\hat L_2$. The $y=+\infty$ limit of equation $\hat L_2\vec\phi=\vec 0$ 
yields $\vec\sigma_t+\lambda\vec\sigma_y=\vec 0$, whose solution is (\ref{t-dep-sigma-chi}a). Analogously, 
\beq
\label{def-pi}
{\pi_{\pm}}_1(\vec x,z,\lambda,t)\equiv {\psi_{\pm}}_1(\vec x,z,\lambda,t), ~~~~ 
{\pi_{\pm}}_2(\vec x,z,\lambda,t)\equiv {\psi_{\pm}}_2(\vec x,z,\lambda,t)-\lambda t
\eeq
are a basis of common analytic eigenfunctions of $\hat L_1$ and $\hat L_2$; therefore 
\beq
{\pi_{\pm}}_1={{\check{\cal K}}_{\pm 1}}(\vec\phi,\lambda),~~~
{\pi_{\pm}}_2={\check{\cal K}_{\pm 2}}(\vec\phi,\lambda), 
\eeq
for some functions ${\check{\cal K}_{\pm 1,2}}$ depending on $(\vec x,z,t)$ only through $\vec\phi$. 
Comparing, at $t=0$, these equations 
with equations (\ref{varphi-psi}), one expresses ${\check{\cal K}_{\pm {1,2}}}$  
in terms of ${{\cal K}_{\pm}}_{1,2}$, obtaining equations (\ref{t-dep-sigma-chi}b).

\subsection{The heavenly reduction}
In the heavenly (Hamiltonian) reduction (\ref{red-u}),(\ref{N=2}), the transformations 
$\vec\xi\to\vec{\cal S}(\vec\xi,\lambda)$, 
$\vec\xi\to\vec{\cal K}_{\pm}(\vec\xi,\lambda)$ are constrained to be canonical:
\beq
\label{constraint-SK1}
\{{\cal S}_1,{\cal S}_2\}_{\vec\xi}=\{{{\cal K}_{\pm}}_1,{{\cal K}_{\pm}}_2\}_{\vec\xi}=1,
\eeq
or, in terms of $\vec\sigma(\xi,y,\lambda)$ and $\vec\chi_{\pm}(\xi,y,\lambda)$:
\beq
\label{constraint-SK2}
\sigma_{1\xi}+\sigma_{2y}+\{\sigma_1,\sigma_2\}_{\vec\xi}=
{\chi_{\pm}}_{1\xi}+{\chi_{\pm}}_{2y}+\{{\chi_{\pm}}_1,{\chi_{\pm}}_2\}_{\vec\xi}=0.
\eeq
To prove it, we recall that the Poisson bracket of the eigenfunctions $\varphi_1$ and $\varphi_2$ is also an eigenfunction:
$\hat L_1\{\varphi_1,\varphi_2\}_{\vec x}=0$.  
Using the asymptotics (\ref{def-varphi}), one infers that $\{\varphi_1,\varphi_2\}_{\vec x}\to 1,~$ at $z\to -\infty$; 
therefore, by uniqueness, $\{\varphi_1,\varphi_2\}_{\vec x}=1$. Evaluating now this Poisson bracket at $z=+\infty$ and using  
(\ref{def-S}), one obtains the constraint (\ref{constraint-SK1}) for $\vec{\cal S}$. We also observe that the 
eigenfunctions $\{{\psi_+}_1,{\psi_+}_2\}_{\vec x}$ and 
$\{{\psi_-}_1,{\psi_-}_2\}_{\vec x}$ are analytic in the upper and lower halves of the $\lambda$-plane and go to $1$ at 
$|\lambda |\to\infty$. 
Since $1$ is also an eigenfunction, by uniqueness they are identically $1$: $\{{\psi_{\pm}}_1,{\psi_{\pm}}_2 \}_{\vec x}=1$. 
Therefore, from the equations:
\beq
\{{\psi_{\pm}}_1,{\psi_{\pm}}_2 \}_{\vec x}=
\{{{\cal K}_{\pm}}_1,{{\cal K}_{\pm}}_2 \}_{(\varphi_1,\varphi_2)}\{\varphi_1,\varphi_2 \}_{\vec x}=1,
\eeq
consequence of (\ref{varphi-psi}), one infers  the constraints (\ref{constraint-SK1}) for $\vec{\cal K}_{\pm}$.
\section{Commuting $\lambda$-families of dynamical systems}
\vskip 5pt
\noindent
It is well-known (see, f.i., \cite{CH}) that linear first order PDEs 
like (\ref{Lax1}) are intimately related to systems of ordinary differential equations describing their 
characteristic curves. The vector fields $\hat L_{1,2}$ (\ref{L1L2bis}) are associated with the following 
two $\lambda$-families of commuting dynamical systems:    
\beq
\label{flow}
\ba{ll}
\hat L_1:   & 
\frac{d\vec x}{dz}=(\lambda,0)+\vec u _1(\vec x,z,t),  \\
~~ \\
\hat L_2:  & 
\frac{d\vec x}{dt}=(0,\lambda)+\vec u_2(\vec x,z,t).
\ea
\eeq
In the heavenly reduction (\ref{red-u}),(\ref{N=2}), they read:
\beq
\label{flow-Ham}
\ba{ll}
\hat L_1:   & 
\frac{d\vec x}{dz}+\{H_1+\lambda y,\vec x \}_{\vec x}=\vec 0,  \\
~~ \\
\hat L_2:  & 
\frac{d\vec x}{dt}+\{H_2-\lambda x,\vec x \}_{\vec x}=\vec 0.
\ea
\eeq
Therefore: {\it the heavenly equation characterizes the commutation, $\forall \lambda$, of  
two $\lambda$-families of Hamiltonian dynamical systems, with Hamiltonians}:
\beq
\tilde H_1=H_1(\vec x,z,t)+\lambda y,~~\tilde H_2=H_2(\vec x,z,t)-\lambda x.
\eeq
 
There is a deep connection between the above IST and the $z$-scattering theory for the commuting flows 
(\ref{flow}) and (\ref{flow-Ham}). Let $\vec\phi(\vec x,z,\lambda,t)$ be the common eigenfunctions of $\hat L_1$ and 
$\hat L_2$ defined in (\ref{def-phi}); then, solving the system 
$\vec\omega=\vec\phi(\vec x,z,\lambda,t)$ with respect to $x$ and $y$ (assuming local invertibility), one obtains the following 
common solution of the commuting flows (\ref{flow}):
\beq
\vec\omega=\vec\phi(\vec x,z,\lambda,t)~~\Leftrightarrow~~\vec x=\vec r(z,t,\lambda,\vec\omega)~\sim~\left(
\ba{c}
\lambda z \\
\lambda t 
\ea
\right)+\vec\omega,~~z\sim -\infty.
\eeq
The $z=+\infty$ limit of the solution $\vec r(z,t,\lambda,\vec\omega)$:
\beq
\vec x=\vec r(z,t,\lambda,\vec\omega)~\sim~\left(
\ba{c}
\lambda z \\
\lambda t 
\ea
\right)+\vec\Omega(\vec\omega,\lambda),~~y\sim +\infty
\eeq
defines the scattering vector $\vec\Delta(\vec\omega)=\vec\Omega(\vec\omega)-\vec\omega$ 
of (\ref{flow}), which is connected to the IST data $\vec{\cal S}$ by inverting the 
system $\vec\omega=\vec{\cal S}(x-\lambda z,y-\lambda t,\lambda)$ with 
respect to $x$ and $y$:
\beq
\vec\omega=\vec{\cal S}(x-\lambda z,y-\lambda t,\lambda,0)~~\Leftrightarrow~~\vec x=
\left(
\ba{c}
\lambda z \\
\lambda t
\ea
\right)+\vec\Omega(\vec\omega,\lambda).
\eeq
In the heavenly reduction, the transformation $\vec\omega \to \vec\Omega(\vec\omega,\lambda)$ is clearly canonical: 
$\{{\Omega}_1,{\Omega}_2\}_{\vec\omega}=1$, and the scattering vector $\vec\Delta$ exhibits the 
following constraint:
\beq
{\Delta_1}_{\omega_1}+{\Delta_2}_{\omega_2}+\{\Delta_1,\Delta_2 \}_{\vec\omega}=0.
\eeq 
The above IST theory allows one to reconstruct, from the scattering vector $\vec\Delta$, the potentials $\vec u_{1,2}$ 
of the dynamical systems (\ref{flow}) and (\ref{flow-Ham}).

\section{Other inverse problems}

Due to the ring property of the space of eigenfunctions, there are other distinguished ways to do the inverse problem. 

\subsection{A nonlinear RH problem}

We begin with a more traditional (nonlinear) Riemann-Hilbert (RH) problem.  
Solving the algebraic system (\ref{varphi-psi})$_-$ with respect to $\vec\varphi$: $\vec\varphi=L(\vec\psi_{-},\lambda)$ 
(assuming local invertibility) and replacing   
this expression in the algebraic system (\ref{varphi-psi})$_+$, one obtains the representation of the analytic eigenfunction 
$\vec\psi_{+}$ in terms of the analytic eigenfunction $\vec\psi_{-}$:
\beq
\label{RH}
\vec\psi_{+}=\vec{\cal R}(\vec\psi_{-},\lambda)=\vec\psi_{-}+\vec R(\vec\psi_{-},\lambda),~~\lambda\in\RR,
\eeq
which defines a {\it vector nonlinear RH problem on the real $\lambda$ axis}. The RH 
data $\vec R$ are therefore constructed from the data $\vec\chi_{\pm}$ by algebraic manipulation. Viceversa, 
given the RH data $\vec R$, one constructs the solutions $\vec\psi_{\pm}$ of the nonlinear RH problem (\ref{RH}) and, 
via the asymptotics (\ref{asympt_psi}), the potential $\vec u_1$. 

As for the other spectral data, one can show that 
the $t$-dependence of $\vec R$ is described by $\vec R(\vec\xi,\lambda,t)=\vec R(\vec\xi-(0,\lambda)t,\lambda,0)$, 
and the reality constraint takes the following form, for $\lambda\in\RR$: 
$\vec{\cal R}(\overline{ \vec{\cal R}(\bar{\vec\zeta},\lambda)},\lambda)=\vec\zeta,~\forall\vec\zeta$. 
At last, the heavenly constraint reads 
$\{{\cal R}_1,{\cal R}_2\}_{\vec\zeta}=1$, or, in terms of $\vec R(\vec\zeta,\lambda)$:
\beq
\label{constraint-R}
R_{1\zeta_1}+R_{2\zeta_2}+\{R_1,R_2\}_{\vec\zeta}=0.
\eeq

\subsection{Linearization of the inverse problem}

It is possible to construct a linear version of the inverse problem of Section 3.3 by 
{\it exponentiating the Jost and analytic eigenfunctions} used so far. Consider the 
following scalar functions:
\beq
\label{def-Phi-Psi}
\Phi(\vec x,z,\lambda,\vec\alpha)\equiv e^{i\vec\alpha\cdot\vec\varphi(\vec x,z,\lambda)},~~~
\Psi_{\pm}(\vec x,z,\lambda,\vec\alpha)\equiv e^{i\vec\alpha\cdot\vec\psi_{\pm}(\vec x,z,\lambda)},~~~~\vec\alpha\in\RR^2.
\eeq
Due to the ring property of the space of eigenfunctions, also $\Phi$ and $\Psi_{\pm}$ 
are eigenfunctions; $\Phi$ is characterized by the asymptotics 
$\Phi\to exp(i\vec\alpha\cdot\vec \xi),~z\to -\infty$, while $\Psi_{\pm}(\vec x,z,\lambda\vec\alpha)$ are analytic 
respectively in the 
upper and lower halves of the $\lambda$ plane, with asymptotics: 
\beq
\label{asympt-Psi}
\Psi_{\pm}=e^{i\vec\alpha\cdot\vec \xi}\left(1+\frac{i\vec\alpha\cdot\vec Q_{\pm}(\vec x,z)}{\lambda}+O(\lambda^{-2})\right).
\eeq

From equations (\ref{def-Phi-Psi}) and (\ref{varphi-psi}) it follows that
\beq
\label{intermediate}
\Psi_{\pm}(\vec x,z,\lambda,\vec\alpha)=\Phi(\vec x,z,\lambda,\vec\alpha)+
e^{i\vec\alpha\cdot\vec\varphi(\vec x,z,\lambda)}
\left(e^{i\vec\alpha\cdot\vec\chi_{\pm}(\vec\varphi(\vec x,z,\lambda),\lambda)}-1\right).
\eeq
The last term of this equality is clearly equal to its anti-Fourier transform:
\beq
\label{anti-Fourier}
e^{i\vec\alpha\cdot\vec\varphi}
\left(e^{i\vec\alpha\cdot\vec\chi_{\pm}(\vec\varphi,\lambda)}-1\right)=
\int_{\RR^2}d\vec\beta K_{\pm}(\vec\alpha,\vec\beta,\lambda)e^{i\vec\beta\cdot\vec\varphi},~~\vec\xi\in\RR^2, 
\eeq
where
\beq
\label{def-K}
\ba{l}
K_{\pm}(\vec\alpha,\vec\beta,\lambda)\equiv \int_{\RR^2}\frac{d\vec \xi}{(2\pi)^2}e^{i(\vec\alpha-\vec\beta)\cdot\vec \xi}
[e^{i\vec\alpha\cdot\vec\chi_{\pm}(\vec\xi,\lambda)}-1].
\ea
\eeq 
Then, using (\ref{def-Phi-Psi}a),  
the last term in the RHS of (\ref{intermediate}) becomes a  linear expression in terms of $\Phi$, and   
equations (\ref{intermediate}) give  
the {\it linear} representations the analytic eigenfunctions $\Psi_{\pm}$ in terms of the Jost eigenfunction $\Phi$:
\beq
\label{Phi-Psi}
\Psi_{\pm}(\vec x,z,\lambda,\vec\alpha)=\Phi(\vec x,z,\lambda,\vec\alpha)+
\int_{\RR^2}d\vec\beta K_{\pm}(\vec\alpha,\vec\beta,\lambda)\Phi(\vec x,z,\lambda,\vec\beta).
\eeq
Multiplying the equations (\ref{Phi-Psi})$_{+}$ and (\ref{Phi-Psi})$_{-}$ by 
$exp(-i\vec\alpha\cdot\vec \xi)$, subtracting $1$, applying respectively $\hat P_{-}$ and $\hat P_{+}$,  
and adding the resulting equations, one obtains the following {\it linear integral equation} for $\Phi$: 
\beq
\label{Phi-int-equ}
\ba{l}
\Phi(\lambda,\vec\alpha)+\frac{1}{2\pi i}\int_{\RR}\frac{d\lambda'}{\lambda'-(\lambda+i\epsilon)}

\int_{\RR^2}d\vec\beta K_-(\vec\alpha,\vec\beta,\lambda')\Phi(\lambda',\vec\beta)
e^{i\alpha_1(\lambda'-\lambda)z} - \\
~~ \\
-\frac{1}{2\pi i}\int_{\RR} \frac{d\lambda'}{\lambda'-(\lambda-i\epsilon)}
\int_{\RR^2}d\vec\beta K_+(\vec\alpha,\vec\beta,\lambda')\Phi(\lambda',\vec\beta)
e^{i\alpha_1(\lambda'-\lambda)z}=e^{i\vec\alpha\cdot\vec\xi}, 
\ea
\eeq
in which we have omitted, for simplicity, the parametric dependence on $(\vec x,z)$ of $\Phi$. 
Once $\Phi$ is reconstructed from (\ref{Phi-int-equ}) and, via (\ref{Phi-Psi}), $\Psi_{\pm}$ are also known, 
the potentials  are reconstructed in the usual way from the asymptotics (\ref{asympt-Psi}) of $\Psi_{\pm}$. 

We finally observe that the reality constraints for the eigenfunctions $\Phi,~\Psi_{\pm}$ and for the data $K_{\pm}$ 
read, for $\lambda\in\RR$,
\beq
\ba{l}
\overline{\Phi(\vec x,z,\lambda,\vec\alpha)}=\Phi(\vec x,z,\lambda,-\vec\alpha),~~
\overline{\Psi_+(\vec x,z,\lambda,\vec\alpha)}=\Psi_-(\vec x,z,\lambda,-\vec\alpha), \\
\overline{K_+(\vec\alpha,\vec\beta,\lambda)}=K_-(-\vec\alpha,-\vec\beta,\lambda),
\ea
\eeq 
while the $t$-evolution of $K_{\pm}$ is given by:
\beq
K_{\pm}(\vec\alpha,\vec\beta,\lambda,t)=K_{\pm}(\vec\alpha,\vec\beta,\lambda,0)e^{i\lambda(\alpha_2-\beta_2)t}.
\eeq
\section{Dressing}
Dressing schemes corresponding to the three inverse problems presented in this paper can always be formulated (see, f.i., 
\cite{BM} for a general treatment).  
Here we restrict our attention to that corresponding to the RH inverse problem   
of Section 5.1, but for the eigenfunctions $\vec\pi_{\pm}$ defined  
in (\ref{def-pi}). The proof, quite standard in the dressing philosophy, is left to the reader.

\vskip 5pt
\noindent
{\it Consider the following nonlinear RH problem on the real $\lambda$-axis}: 
\beq
\label{dressing-pi}
\vec\pi_{+}=\vec\pi_{-}+\vec{\check{R}}(\vec\pi_{-},\lambda)\equiv\vec{\check{\cal R}}(\vec\pi_{-},\lambda),~~\lambda\in\RR,
\eeq
{\it for the functions $\vec\pi_{+}(\vec x,z,t,\lambda)$ and $\vec\pi_{-}(\vec x,z,t,\lambda)$,   
analytic respectively in the upper and lower halves of the complex plane $\lambda$, with asymptotics:}
\beq
\label{asympt_pi}
\vec\pi_{\pm}(\vec x,z,\lambda)=\left(
\ba{c}
x-\lambda z \\
y-\lambda t
\ea
\right)+\frac{\vec Q_{\pm}(\vec x,z,t)}{\lambda}+O(\lambda^{-2}).
~~~|\lambda|>>1,
\eeq  
{\it where the vector $\vec{\check{R}}(\vec\pi_{-},\lambda)$ depends on $(\vec x,z,t)$ only through $\vec\pi_{-}$. 
Assume also the unique solvability of the RH problem (\ref{dressing-pi}) and of its linearized version}
\beq
\label{lin-dressing-pi}
\ba{l}
\vec\nu_{+}=\vec\nu_{-}+\rho(\vec\pi_{-},\lambda)\vec\nu_-,~~\lambda\in\RR, \\
(\rho)_{ij}(\vec\zeta,\lambda)\equiv \frac{\partial {\check R}_i(\vec\zeta,\lambda)}{\partial\zeta_j}. 
\ea
\eeq
{\it Then $\vec\pi_{\pm}$ are solutions of $\hat L_1\vec\pi_{\pm}=\hat L_2\vec\pi_{\pm}=\vec 0$, where 
$\hat L_1,\hat L_2$ are defined in (\ref{L1L2bis}) with:}
\beq
\vec u_1(\vec x,z,t)=-{\vec Q_{\pm x}},~~\vec u_2(\vec x,z,t)=-{\vec Q_{\pm y}}.
\eeq
{\it It follows that the potentials $\vec u_{1,2}$ are solutions of the nonlinear system of PDEs (\ref{quasilin-U-bis}). 
If, in addition, the spectral data  satisfy the Hamiltonian constraint}
\beq
\check{R}_{1\zeta_1}+\check{R}_{2\zeta_2}+\{\check{R}_1,\check{R}_2\}_{\vec\zeta}=0,
\eeq 
{\it then the potentials $\vec u_{1,2}$ satisfy the heavenly equation (\ref{heavenly1}). At last, the reality constraints }
\beq
\vec{\check{\cal R}}(\overline{ \vec{\check{\cal R}}(\bar{\vec\zeta},\lambda)},\lambda)=
\vec\zeta,~~\forall\vec\zeta,~~\lambda\in\RR
\eeq
{\it imply that $\overline{\vec\pi_+}=\vec\pi_-$, and the reality of the potentials: $\vec u_1,\vec u_2\in\RR^2$}.

\vskip 10pt
\noindent
{\bf Acknowledgements}. The visit of SVM to Rome was supported by the INFN grant 2005 and by the RFBR grant 04-01-00508.


\begin{thebibliography}{9}

\bibitem{Pleb} 
J. F. Plebanski, ``Some solutions of complex Einstein equations'', J. Math. Phys. {\bf 16}, 2395-2402 (1975).

\bibitem{YM}
C. N. Yang and R. L. Mills, ``Conservation of isotopic spin and isotopic gauge invariance'', 
Phys. Rev {\bf 96}, 191-195 (1954).
 
\bibitem{ZS} 
V. E. Zakharov and A. B. Shabat, Functional Anal. Appl. {\bf 13}, 166-174 (1979).

\bibitem{ZMNP} V. E. Zakharov, S. V. Manakov, S. P. Novikov and L. P. Pitaevsky, {\it Theory of solitons}, 
Plenum Press, New York, 1984. 

\bibitem{AC} M. J. Ablowitz and P. A. Clarkson, 
{\it Solitons, nonlinear evolution equations and Inverse Scattering}, London Math. Society Lecture Note 
Series, vol. 194, Cambridge University Press, Cambridge (1991). 

\bibitem{CMN} S. Chakravarty, L. Mason and E. T. Newman, ``Canonical structures on anti-selfdual 
four manifolds and the diffeomorphism group'', J. Math. Phys. {\bf 32}, 1458-1464, (1991).
 
\bibitem{DuMa} M. Dunajski and L. J. Mason, ``Hyper-K\"ahler hierachies and their twistor theory'', 
Comm. Math. Phys. {\bf 213}, 641-672 (2000). ``Twistor theory of hyper-K\"ahler metrics with 
hidden symmetries'', J. Math. Phys., {\bf 44}, 3430-3454 (2003). 

\bibitem{DMT} M. Dunajski, L. J. Mason and P. Todd, ``Einstein-Weyl geometry, the dKP equation and twistor 
theory'', J. Geom. Phys. {\bf 37} 63-93 (2001).
 
\bibitem{NNS}
F. Neyzi, Y. Nutku and M. B. Sheftel, ``A multi-hamiltonian structure of the Plebanski's second heavenly equation''; 
J. Phys. A: Math. Gen. {\bf 38} 8473- (2005).

\bibitem{MM} 
M. Manas and L. Martinez Alonso, ``A hodograph transformation which applies to the heavenly equation''; 
arXiv:nlin.SI/0209050 v1.

\bibitem{BK} 
L. V. Bogdanov and B. G. Konopelchenko, ``On the $\bar\partial$-dressing method applicable to heavenly equation''; 
Phys. Lett. A {\bf 345} (2005) 137-143; ``On the heavenly equation hierarchy and its reductions'', arXiv:nlin.SI/0512074. 

\bibitem{MS} S. V. Manakov and P. M. Santini: ``Inverse scattering problem for 
vector fields and the heavenly equation''; http://arXiv:nlin.SI/0512043.

\bibitem{MZ}
S. V. Manakov and V. E. Zakharov, ``Three-dimensional model of relativistic-invariant field theory, 
integrable by the inverse scattering transform''; Letters in Mathematical Physics {\bf 5}, 247-253 (1981). 

\bibitem{Manakov1} S. V. Manakov, ``The inverse scattering transform for the time - dependent 
Schr\"odinger operator and Kadomtsev-Petviashvili equation'', Physica {\bf 3D}, 420-427 (1981).

\bibitem{CH}
R. Courant and D. Hilbert, {\it Methods of Mathematical Physics; Vol. II: Partial Differential Equations, by R. Courant}, 
Interscience Publishers, J. Wiley and sons, New York, 1962.

\bibitem{BM}
L.V.Bogdanov and S.V.Manakov, J.Phys.A:Math.Gen. {\bf 21}, L537 (1988)

\end{thebibliography}
\end{document}